# High precision analysis of unidirectional pedestrian flow within the Hermes Project

Jun Zhang[a,*], Wolfram Klingsch[a], Armin Seyfried[b,c]

[a]*Institute for Building Material Technology and Fire Safety Science, Wuppertal University, Pauluskirchstrasse 11, 42285,Wuppertal, Germany*
[b]*Computersimualtion for Fire Safety and Evacuation, Wuppertal University, Pauluskirchstrasse 11, 42285,Wuppertal, Germany*
[c]*Jülich Supercomupting Center, Forschungszentrum Jülich Gmbh, 52425 Jülich, Germany*

**Abstract**

The aim of the Hermes project is the development of a computer simulation based on evacuation assistant to support security services in case of emergency in complex buildings and thus to improve safety at mass events. One goal of the project is to build and to calibrate models for pedestrian dynamics specifically designed for forecasting the emergency egress of large crowds faster than real-time. In this contribution, we give an overview of the project and its experimental results of unidirectional flow in a corridor. Trajectories of all pedestrians are analyzed with different measurement methods. The data will be used for model calibration.

## 1. Introduction

With increasing urbanization, more and more emergency incidents such as fire, terrorist attacks etc occur. Meanwhile, multifunctional building structures in combination with a wide range of large-scale public events present new challenges for the quality of security concepts. Prescriptive construction and planning regulations ensure in general that in case of an emergency every related person is able to leave the danger zone quickly by specifying e.g. minimal width and maximal length of escape routes. In the event of loss of rescue routes due to fire or other risks, however, dangerously high crowd densities and bottleneck effects can occur. To prevent such critical situations optimal crowd management needs accurate information about the current status. Usually in complex buildings the decision makers miss information like the number of people in the danger zone, how the loss of escape routes influences the evacuation time or where dangerous congestions with long waiting times will occur in the course of the evacuation.

The evacuation assistant, developed by the Hermes project and outlined in this contribution will close this gap and support the decision makers in actual danger situations, to decide a successful evacuations strategy and to deploy the security staff optimally. The ESPRIT arena in Düsseldorf (Germany) provides a venue for testing the evacuation assistant. The example of this multifunctional arena with a capacity of 60,000 visitors will show how crowds of people at large events can be guided – also considering the current risk situation. A test system of the assistant will be installed in 2011.

Fig. 1 shows the layout of the evacuation assistant. Using automated counting of persons at entrances and doors the present position of people in the building is delivered to the decision makers and the simulation core. The safety and security management system provides information about quality of escape routes like blocked due to smoke, locked doors or other dangers. Using the actual data about the distribution of people and the availability of rescue routes, a parallel computer will generate faster than real-time simulations to predict the movement of all people over the next 15 minutes and update the prediction at one-minute intervals. The simulation predicts results like potential dangerous congestion areas and evacuation times. Moreover a macroscopic network model will calculate the optimal guiding of occupants on the available routes. A communication module will provide this information to the emergency teams on site. Various universities, industrial partners and end users cooperate in this project. For an overview we refer to [1].

---

* Corresponding author. Tel.: +49 202 439 4241; fax: +49 202 82560.
*E-mail address*: jun.zhang@uni-wuppertal.de

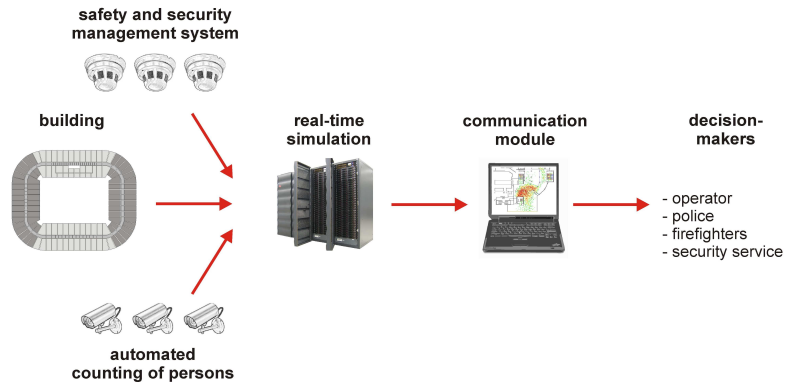

Fig. 1. Schematic diagram of the evacuation assistant.

Different modeling strategies are combined to obtain reliable predictions from the simulations and achieve an optimal performance. One approach uses cellular automata (CA), the other a spatially continuous forced-based model. Both approaches have advantages and disadvantages and we refer to [2] for a deeper discussion. A crucial point in the application of these models to security sensitive tasks within the Hermes project is their quantitative verification and calibration.

Regarding the reliability of simulation results based on these models, we are still at the beginning, see e.g. [3]. This deficiency is due to the contradictory experimental data base for model testing [2]. One of the most important characteristics of pedestrian dynamics is the fundamental diagram giving the relationship between pedestrian flow and density. Several researchers, in particular Fruin and Pauls[4], Predtechenskii and Milinskii[5], Weidmann[6], Helbing[7] have collected detailed information about occupant density and velocity. However, there exists considerable disagreement among these empirical data. As can be seen in the comparison of reference [8], the density ρ0 where the velocity approaches zero due to overcrowding ranges from 3.8 $m^{-2}$ to 10 $m^{-2}$, while the density $ρ_c$ where the flow reaches maximum ranges from 1.75 $m^{-2}$ to 7 $m^{-2}$. Several explanations for these have been proposed, including cultural factors, differences between unidirectional and multidirectional flow [9, 10] and even the different measurement methods used [8]. Facing such questions, a series of well controlled laboratory experiments were carried out within the Hermes project to calibrate and test pedestrian movement models. In the following, we introduce the experiments of unidirectional pedestrian flow through a straight corridor and present preliminary results.

## 2. Experiment Description

The experiments were performed in hall 2 of the exhibition center of Düsseldorf Fair Germany May 2009. Up to 350 probands participated in the experiments. Each participant received 50 € per day, most of them were students (43% female and 57% male). The age and height of the probands were 25 (± 5.7) years-old and 1.76 (± 0.09) *m* respectively. For 42 participants we measured the free velocity $v_0$= 1.55 (± 0.18) *m/s*.

Fig. 2(a) shows the sketch of the experiment setup. We performed 28 runs (see Table 1) in straight corridors (uo) with widths $b_{cor}$ of 1.8 *m*, 2.4 *m* and 3.0 *m*. At the beginning, the participants were held within a waiting area. When the experiment begins, they pass through a passage, which is 4 meters away from the corridor. To alter the pedestrian density in the corridor, the widths of the passage $b_{entrance}$ and the exit $b_{exit}$ were changed in each run, see Table 1 for details. When a pedestrian leaves through the exit, he or she returns to the waiting area to wait for the next run. The experiments were recorded by two cameras mounted at a rack of the ceiling of the hall. To cover the overall region, the left and right parts of the corridor are recorded separately by the two cameras. For analysis, the pedestrian trajectories were extracted using the software *PeTrack* [11], which is able to extract trajectory data automatically from video recordings. Finally, the trajectory data from camera 1 and camera 2 were corrected manually and combined. Fig. 2(b) shows the trajectories of the head of every pedestrian in two runs of the experiment. From these trajectories, pedestrian characteristics including flow, density, and velocity are determined.

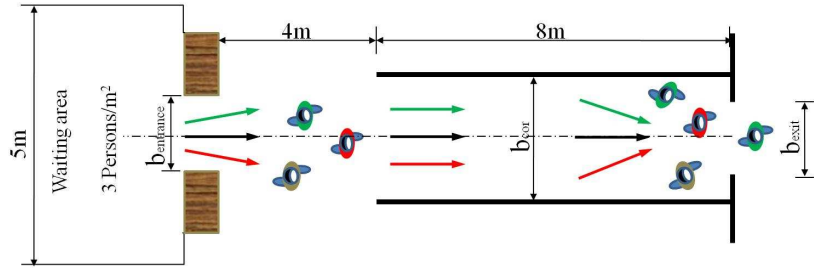

(a) The sketch of the experiment uo

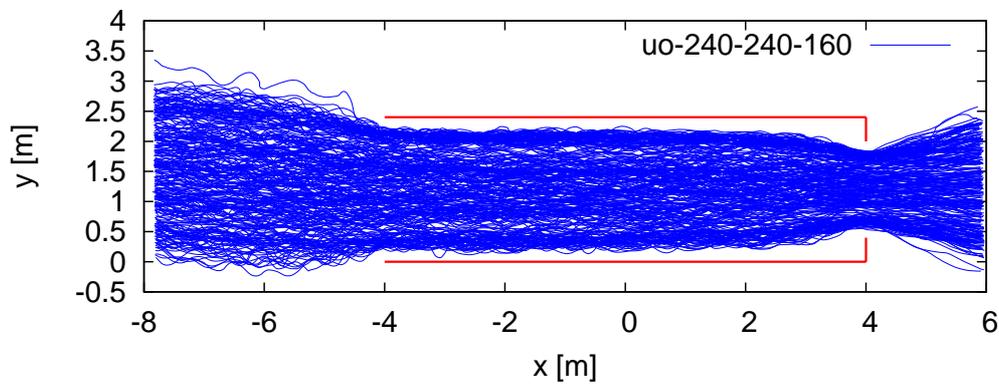

(b) The trajectories for each pedestrian extracted by PeTrack in the experiment uo-240-240-160

Fig. 2. The sketch and the trajectory data of the experiment uo

## 3. Methodology and Results

### 3.1. Measurement Methods

From vehicular traffic it is known that different measurement methods may lead to deviations in the fundamental diagram [12, 13]. In our analysis, two measurement methods are used to calculate the basic quantities of flow, density and velocity. We will analyze which method leads to the smallest fluctuations.

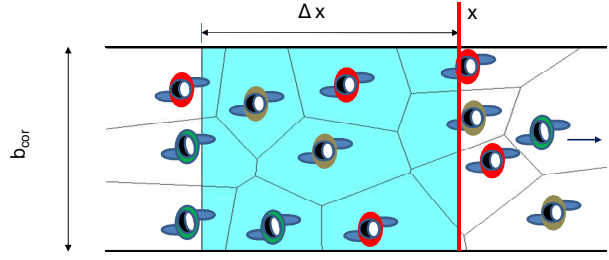

Fig.3. Illustration of different measurement methods.

- Classical Method

The classical method is based on a mean value over space. In this method, a segment Δx in the corridor is taken as measurement area (as shown in Fig. 3). The density $<\rho>_{\Delta x}$ is defined as the number of pedestrians divided by the measurement area:

$$<\rho>_{\Delta x} = \frac{N}{b_{cor} \cdot \Delta x} \quad (1)$$

While the mean velocity $<v>_{\Delta x}$ is the average of the instantaneous velocity $v_i(t)$, which is defined as the movement distance in a small time interval $\Delta t$, for all pedestrians in the measurement area at time $t$.

$$<v>_{\Delta x} = \frac{1}{N}\sum_{i=1}^{N} v_i(t) \quad \text{with} \quad v_i(t) = \frac{x_{t+\Delta t/2} - x_{t-\Delta t/2}}{\Delta t} \quad (2)$$

- Voronoi Method

This method is based upon the Voronoi diagram and was introduced in [14]. The Voronoi diagram is defined by Georgy Voronoi and is published in [15]. At any point in time the pedestrian can be represented as a point $x_i$. The Voronoi diagram is obtained from these points. The Voronoi cell area $A_i$ for each person i consist of all points closer to point $x_i$ than to any other point $x_j$ and allow a unique decomposition of the space (see Fig. 3). The density distribution $\rho_{xy}$ can be defined as:

$$\rho_{xy} = \begin{cases} 1/A_i: & (x,y) \in A_i \\ 0: & otherwise \end{cases} \quad (3)$$

The Voronoi density $<\rho>_v$ for the measurement area is defined as:

$$<\rho>_v = \frac{\iint \rho_{xy} dxdy}{b_{cor} \cdot \Delta x} \quad (4)$$

Similarly, the Voronoi velocity $<v>_v$ can be defined as equation (5).

$$<v>_v = \frac{\iint v_{xy} dxdy}{b_{cor} \cdot \Delta x} \quad \text{with} \quad v_{xy} = \frac{v_i(t)}{A_i(t)} \quad (5)$$

Where $v_i(t)$ is the instantaneous velocity of each pedestrian.

*3.2. Results*

For the measurement area, we choose a rectangle with length 2 *m* from x = -2 *m* to x = 0 *m* and width of $b_{cor}$. We calculate density and velocity every frame with a frame rate of 16 *fps*. The time interval $\Delta t$ = 10 frames is chosen to calculate the instantaneous velocities. All data for diagram below are obtained from trajectories. To determine the fundamental diagram, only data under stationary conditions for each run are selected by analyzing the time series of density and velocity.

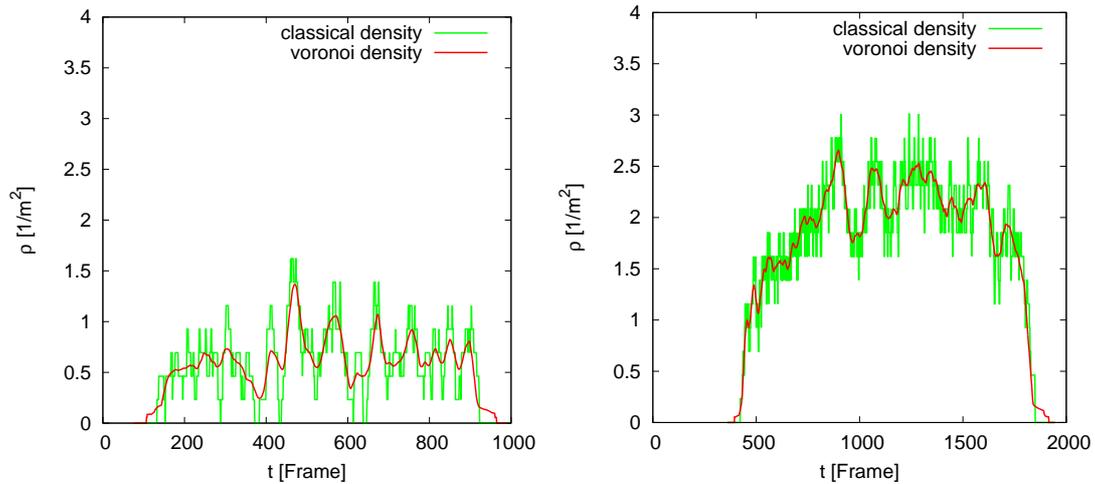

(a) uo-095-240-240  (b) uo-240-240-160

Fig. 4 Time series of the density for different methods

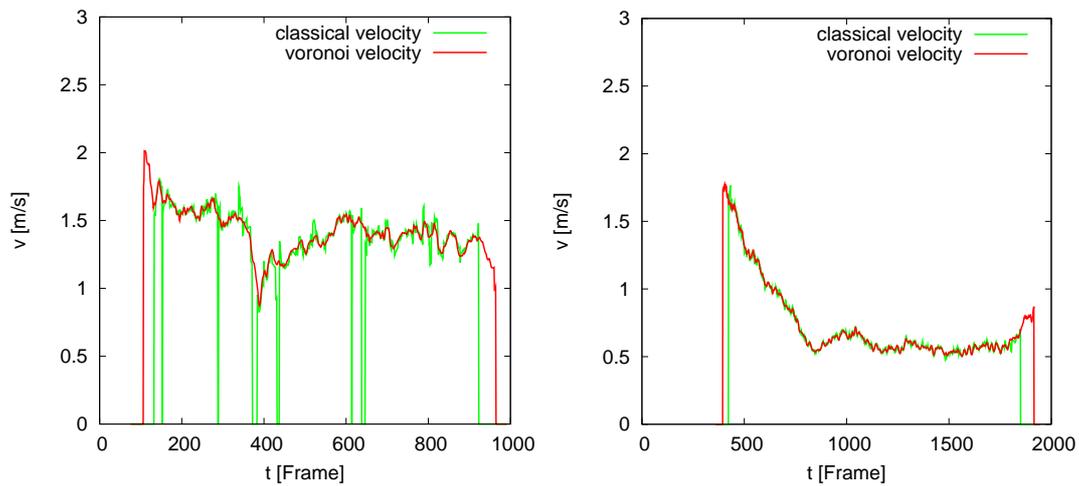

(a) uo-095-240-240  (b) uo-240-240-160

Fig. 5. Time series of the velocity for different methods

Fig. 4 and Fig. 5 show the time sequence of density and velocity gained by the classical method and the Voronoi method respectively. Fig. 4 shows that the Voronoi density is much more stable in time, while the classical density fluctuates largely especially at low densities (a). The Voronoi method is also suitable for the velocity, see Fig. 5. Both classical velocity and Voronoi velocity show only small fluctuations in comparison to density. This is because the changes over time of the instantaneous velocities of pedestrians in the measurement area are small. Due to the small fluctuation of the Voronoi velocity and density, stationary states could be determined easily.

Fig. 6 shows the relationship between density and velocity obtained from these two methods. The trends for the fundamental diagrams gained by different methods are similar. However, the data from the classical method fluctuates more than that from Voronoi method. The discreteness of the density values depends on the dimension of the meassurement area. The interval between two density values is determined by the inverse of the area of the measurement area. Thus the measurement area should not be too small when using this method.

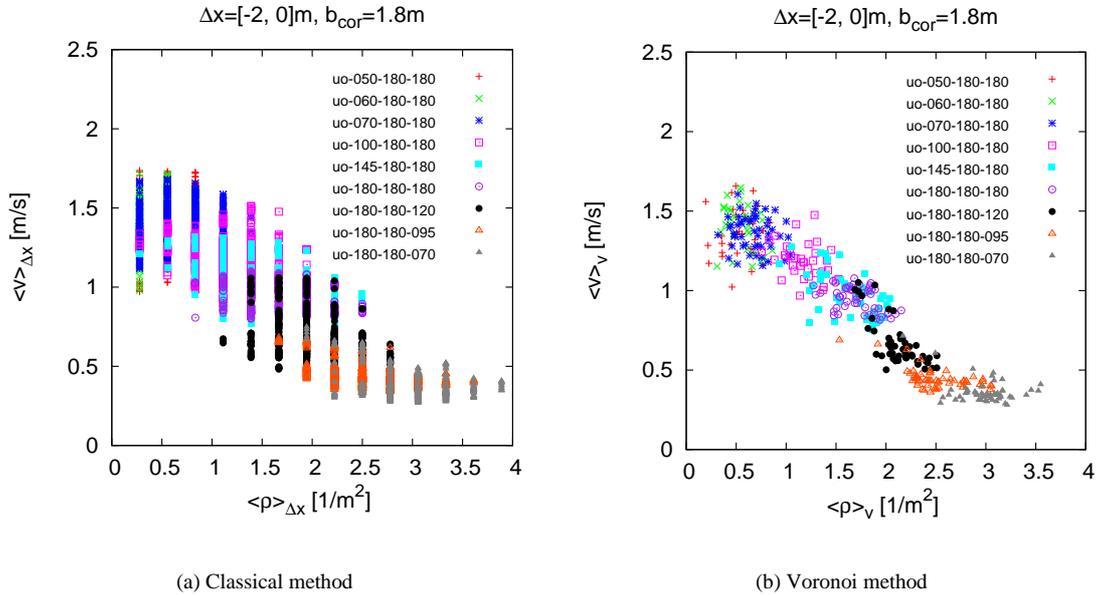

(a) Classical method  (b) Voronoi method
Fig. 6. The fundamental diagram, relationship between density and velocity, measured with the same set of trajectories but with different methods.

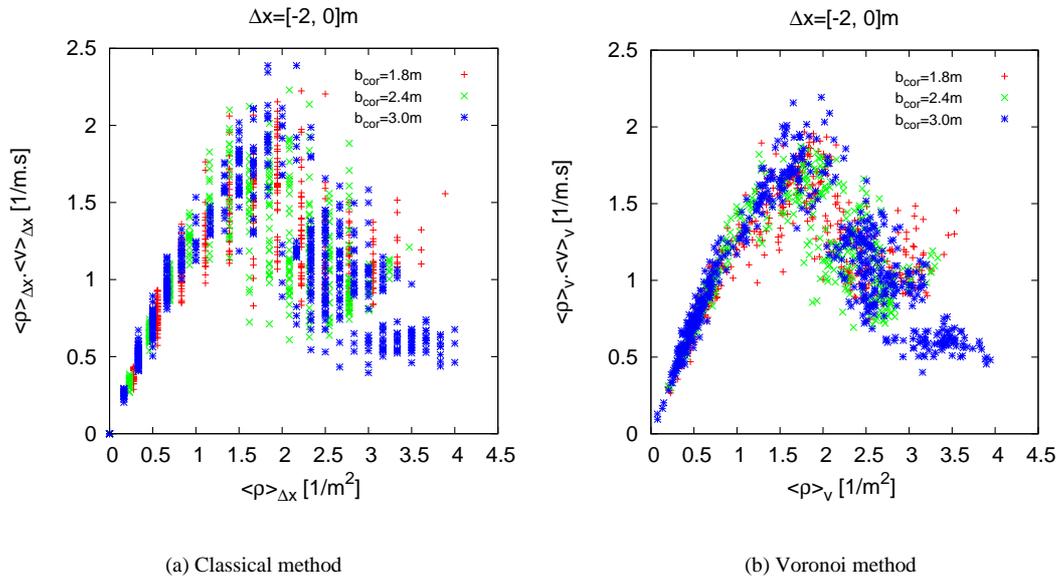

(a) Classical method  (b) Voronoi method
Fig. 7. Comparison of the fundamental diagram for different corridor width

We also analyzed the experiments with corridor width of 2.4 $m$ and 3.0 $m$ using the Voronoi method. Fig. 7 shows the comparison of the fundamental diagrams for both methods of different corridor widths. The relationship between density and specific flow $J_s$ are compared with equation $J_s = \rho v$ used to calculate the specific flow. The fundamental diagrams for these three widths agree very well. This result supports the assumption that flow-density relations for the same type of facility could be unified into one diagram for the specific flow. This result agrees with Hankin's findings [16]. He found that above a certain minimum of about 1.22 *m* the maximum flow in subways is directly proportional to the width of the corridor, while multiples of shoulder widths becomes important for corridor width below 1.22 *m*.

## 4. Summary

A series of controlled laboratory pedestrian experiments were performed by the Hermes project to calibrate and test pedestrian models. In this study, experiments through straight corridors are examined. The combination of modern video equipment with new methods for extracting relevant data allows an unprecedented view into pedestrian behaviour. The trajectories extracted from the video recordings and two different measurement methods were used to analyze the fundamental diagram in straight corridors. The fundamental diagrams obtained by different methods show the same trend but with different precision. It was found that the fundamental diagrams for the same type of facility can be unified into one diagram for specific flow.

## Acknowledgements

The authors thank to the financial support from the program of China Scholarships Council. This work has been performed within the program "Research for Civil Security" in the field "Protecting and Saving Human Life" funded by the German Government, Federal Ministry of Education and Research (BMBF).

# Appendix A.

Table 1. The related parameters in straight corridor experiment

| Experiment index | Name | $b_{entrance}$ [m] | $b_{cor}$ [m] | $b_{exit}$ [m] | N |
|---|---|---|---|---|---|
| 1 | uo-050-180-180 | 0.50 | 1.80 | 1.80 | 61 |
| 2 | uo-060-180-180 | 0.60 | 1.80 | 1.80 | 66 |
| 3 | uo-070-180-180 | 0.70 | 1.80 | 1.80 | 111 |
| 4 | uo-100-180-180 | 1.00 | 1.80 | 1.80 | 121 |
| 5 | uo-145-180-180 | 1.45 | 1.80 | 1.80 | 175 |
| 6 | uo-180-180-180 | 1.80 | 1.80 | 1.80 | 220 |
| 7 | uo-180-180-120 | 1.80 | 1.80 | 1.20 | 170 |
| 8 | uo-180-180-095 | 1.80 | 1.80 | 0.95 | 159 |
| 9 | uo-180-180-070 | 1.80 | 1.80 | 0.70 | 148 |
| 10 | uo-065-240-240 | 0.65 | 2.40 | 2.40 | 70 |
| 11 | uo-080-240-240 | 0.80 | 2.40 | 2.40 | 118 |
| 12 | uo-095-240-240 | 0.95 | 2.40 | 2.40 | 108 |
| 13 | uo-145-240-240 | 1.45 | 2.40 | 2.40 | 155 |
| 14 | uo-190-240-240 | 1.90 | 2.40 | 2.40 | 218 |
| 15 | uo-240-240-240 | 2.40 | 2.40 | 2.40 | 246 |
| 16 | uo-240-240-160 | 2.40 | 2.40 | 1.60 | 276 |
| 17 | uo-240-240-130 | 2.40 | 2.40 | 1.30 | 247 |
| 18 | uo-240-240-100 | 2.40 | 2.40 | 1.00 | 254 |
| 19 | uo-080-300-300 | 0.80 | 3.00 | 3.00 | 119 |
| 20 | uo-100-300-300 | 1.00 | 3.00 | 3.00 | 100 |
| 21 | uo-120-300-300 | 1.20 | 3.00 | 3.00 | 163 |
| 22 | uo-180-300-300 | 1.80 | 3.00 | 3.00 | 208 |
| 23 | uo-240-300-300 | 2.4 | 3.00 | 3.00 | 296 |
| 24 | uo-300-300-300 | 3.00 | 3.00 | 3.00 | 349 |
| 25 | uo-300-300-200 | 3.00 | 3.00 | 2.00 | 351 |
| 26 | uo-300-300-160 | 3.00 | 3.00 | 1.60 | 349 |
| 27 | uo-300-300-120 | 3.00 | 3.00 | 1.20 | 348 |
| 28 | uo-300-300-080 | 3.00 | 3.00 | 0.80 | 270 |